\documentclass[aps,prb,twocolumn,superscriptaddress,showpacs,floatfix]{revtex4-1}
\usepackage{graphicx}
\usepackage{amssymb}

\begin{document}

\title{Signatures of quantum criticality in the thermopower of Ba(Fe$_{1-x}$Co$_{x}$)$_{2}$As$_{2}$}

\author{S. Arsenijevi\'c}
\affiliation{Institute of Condensed Matter Physics, Swiss Federal Institute of Technology, EPFL, CH-1015 Lausanne, Switzerland}
\affiliation{Laboratoire National des Champs Magn\'{e}tiques Intenses, LNCMI-CNRS, F-38042 Grenoble, France}

\author{H. Hodovanets}
\affiliation{Ames Laboratory, U.S. DOE, and Department of Physics and Astronomy, Iowa State University, Ames, Iowa 50011, USA}

\author{R. Ga\'{a}l}
\affiliation{Institute of Condensed Matter Physics, Swiss Federal Institute of Technology, EPFL, CH-1015 Lausanne, Switzerland}

\author{L. Forr\'o}
\affiliation{Institute of Condensed Matter Physics, Swiss Federal Institute of Technology, EPFL, CH-1015 Lausanne, Switzerland}

\author{S. L. Bud'ko}
\affiliation{Ames Laboratory, U.S. DOE, and Department of Physics and Astronomy, Iowa State University, Ames, Iowa 50011, USA}

\author{P. C. Canfield}
\affiliation{Ames Laboratory, U.S. DOE, and Department of Physics and Astronomy, Iowa State University, Ames, Iowa 50011, USA}

%\date{\today}

\begin{abstract}
We demonstrate that the thermopower ($S$) can be used to probe the spin fluctuations (SFs) in proximity to the quantum critical point (QCP) in Fe-based superconductors. The sensitivity of $S$ to the entropy of charge carriers allows us to observe an increase of $S/T$ in Ba(Fe$_{1-x}$Co$_x$)$_2$As$_2$ close to the spin-density-wave (SDW) QCP. This behavior is due to the coupling of low-energy conduction electrons to two-dimensional SFs, similar to heavy-fermion systems. The low-temperature enhancement of $S/T$ in the Co substitution range 0.02 $< x <$ 0.1 is bordered by two Lifshitz transitions, and it corresponds to the superconducting region, where a similarity between the electron and non-reconstructed hole pockets exists. The maximal $S/T$ is observed in proximity to the commensurate-to-incommensurate SDW transition, for critical $ x_c\approx$ 0.05, close to the highest superconducting $T_c$. This analysis indicates that low-$T$ thermopower is influenced by critical spin fluctuations which are important for the superconducting mechanism.
\end{abstract}


%\keywords{}

\maketitle

\section{Introduction}

The physical properties of matter in the vicinity of a quantum critical point (QCP) have been the focus of interest since the discovery of unconventional superconductivity~\cite{Bednorz,TailleferARCMP} and heavy-fermion systems~\cite{Lohneysen,GegenwartQCP}. The discovery of superconductivity (SC) in Fe-based materials (FeSC) and the presence of a spin-density-wave (SDW) state motivated discussions about the interplay of magnetism, structure, and superconductivity which coexists with the QCP in the phase diagram of FeSC \cite{Hosono,DaiCriticality}. In FeSC, the structural, tetragonal-to-orthorhombic, transition is coupled to the paramagnetic-to-antiferromagnetic transition~\cite{HuangPRL}. This behavior can be realized through nematic order which emerges from the coexistence of magnetic fluctuations and frustration \cite{XuSachdev,FernandesNematic,FangNematic,KasaharaNematic,Chunematic}. It explains the proximity in temperature of the structural ($T_{\rm S}$) and magnetic ($T_{\rm SDW}$) transitions throughout the phase diagram of doped iron-pnictides~\cite{Nandi}. The observed anisotropy of the in-plane resistivity is in agreement with the nematic scenario of anisotropic electronic states originating from the scattering by impurities and critical spin fluctuations (SFs)~\cite{FernandesNematic,ChuAniso,Tanatar,Eremin}. The study of magnetic fluctuations are important because it is believed they are responsible for the SC pairing~\cite{Scalapino,MiyakeSF}.
% MonthouxSC

The thermoelectric power ($S$) is sensitive to the derivative of the density of electronic states and the change in the relaxation time at the Fermi surface (FS). It can be interpreted as the entropy per charge carrier \cite{Chaikin,Zlatic}. $S$ can be used to detect deviations from the Landau Fermi-liquid (FL) picture, $i.e.$ in heavy-fermion compounds. There, the enhanced scattering by critical spin fluctuations (SFs) close to the antiferromagnetic (AF) quantum critical point leads to an increase of electronic entropy and, consequently, to increases of thermopower and electronic specific heat ($C_e$) \cite{PaulKotliar}. The increase of entropy and $C_e$ upon entering the nematic phase in the vicinity of the quantum critical phase was shown in the example of $\rm Sr_3Ru_2O_7$ \cite{RostEntropy}. In this paper, we observe quantum critical behavior by thermopower in the phase diagram of the prototypical Fe-based superconductor Ba(Fe$_{1-x}$Co$_x$)$_2$As$_2$ (BFCA).

\section{Quantum criticality and thermopower}

The variations of thermopower $S/T$ have been used to characterize the nature of the QCP in non-Fermi-liquid (NFL) heavy-fermion compounds \cite{KimPepin}. In the case of a spin-density-wave criticality, the $S/T$ is roughly symmetric around the QCP. Also, it was shown that $S/T$ near the magnetic quantum critical point has a variation similar to $C_e/T$ \cite{PaulKotliar}. The low-energy quasi-two-dimensional (2D) spin fluctuations with a 2D ordering wave vector and a three-dimensional (3D) Fermi surface lead to ``hot'' regions (with a high scattering rate) on the Fermi surface \cite{RoschNFL}. The electrons are strongly renormalized in these regions because of the enhanced scattering on nearly critical spin fluctuations. This leads to the following expression (taken from Ref. \cite{PaulKotliar}) for the specific heat or entropy per particle:
\begin{equation}
C_e \propto \mathcal{N}(0) T \frac{g_0^2}{\epsilon_F \omega_S} \ln(\omega_S/\delta).
\end{equation}
% \frac{C_e}{N}
Here, $\mathcal{N}(0)$ is the density of states at the Fermi energy $\epsilon_F$, and $g_0^2$ is the bare coupling between the electrons and spin fluctuations. The energy of the spin fluctuations is given by $\omega_S$, where $\omega_S \sim W$, the bandwidth of the conduction electrons, while $\delta$ is the mass of the SF and it measures the deviation from the QCP. The logarithmic $T$ dependence of specific heat is different from the Fermi-liquid behavior in which $C_e \propto T$. Analogously, according to Ref.~\cite{PaulKotliar}, the expression for thermopower based on critical 2D SFs is
\begin{equation}\label{eq:SdT}
\frac{S}{T} \propto \frac{1}{e} (\frac{g_0^2 \mathcal{N}'(0)}{\epsilon_F \omega_S \mathcal{N}(0)}) \ln(\omega_S/\delta).
\end{equation}
One can write $\delta$ as $\delta = \Gamma(p - p_c) + T$, where $\Gamma$ is an energy parameter and $p$ is an experimental parameter (doping, pressure or magnetic field) that can be tuned to the critical value $p_c$. This means that the QCP can be approached by changing the temperature or other parameters in the system. In the former case, when $T > \Gamma(p - p_c)$, $S$ in proximity to QCP has a dependence $S/T \propto \ln(1/T)$, qualitatively different from the FL behavior $S/T \propto$ const. \cite{PaulKotliar}

The NFL divergent behavior of $S/T$ close to the antiferromagnetic SDW QCP is observed in several unconventional superconductors, among others: heavy-fermion $\rm Ce_2PdIn_8$\cite{MatusiakQCP}, cuprate high-$T_c$ superconductor La$_{1.6-x}$Nd$_{0.4}$Sr$_{x}$CuO$_4$ \cite{Daou}, and hole-doped Fe pnictides Ba$_{1-x}$K$_x$Fe$_2$As$_2$, Sr$_{1-x}$K$_x$Fe$_2$As$_2$ \cite{GoochPRB,Gooch}, and Eu$_{1-x}$K$_x$Fe$_2$As$_2$ \cite{Maiwald}. The difference between these compounds is the energy defined by the temperature below which the critical behavior is observed and SC emerges, which is smaller in heavy-fermion and larger in high-$T_c$ SC. Another sign of quantum critical behavior is the $T$-linear resistivity $\rho(T)$ driven by anomalous scattering on spin fluctuations, for the critical value of doping, which was reported in all of the aforementioned compounds \cite{Daou,GoochPRB,Gooch,DongPRX}. Also, the critical behavior of $\rho(T)$ corresponds to the highest SC $T_c$, thus supporting the SF-driven SC scenario \cite{TailleferARCMP}. The highest energy-range of criticality is observed in La$_{2-x}$Sr$_x$CuO$_4$ cuprates, where the linear $T$ dependence of $\rho$ extends up to 1100 K \cite{Gurvitch}. In both cuprates and Fe pnictides, this anomalous behavior is observed only in a narrow doping range, for a critical value of doping \cite{GoochPRB,AndoPRL,Kasahara}.

\section{\emph{S/T} of BFCA -- Quantum criticality and thermopower}

Here we focus on the $S/T$ in the low-$T$ region that shows anomalous behavior in Co substituted, electron-doped BFCA.
SDW long-range AF order at $T_{\rm SDW}$ is defined by a commensurate propagation vector which is the nesting vector between the hole and electron pockets on the Fermi surface \cite{TerashimaNesting}. The $T_{\rm SDW}$ occurs at lower temperature than the $T_S$ transition \cite{KimCanfield} and in the SDW phase the FS is reconstructed~\cite{BFCALifshitz}. With Co substitution, the structural and SDW transitions are suppressed and increasingly separated, and the FS undergoes a Lifshitz transition above $x\approx$ 0.02 \cite{BFCALifshitz,BFCAFStopo}. It is a topological change of the FS, and the first one occurs when the reconstructed hole pocket disappears below the Fermi level, giving way to the electron pocket at the Brillouin zone $X$ corner (LT1). A similarity in the size and shape of this electron $X$ pocket and the hole $\Gamma$ pocket in the zone center exists in the 0.02 $< x <$ 0.1 range \cite{BFCAFStopo}. This feature enhances interband scattering which is important for superconducting pairing \cite{TerashimaNesting,Mazin,KurokiSC,Neupane}. The low-energy spin resonance observed in the SC phase by inelastic neutron scattering at the same nesting vector supports the picture of a SC pairing mechanism mediated by spin fluctuations \cite{Christianson}. Also, nuclear magnetic resonance links the strength of AF SFs and SC $T_c$, when the SDW order is almost suppressed \cite{NingPRL}. In the same region, for $x \approx $ 6\%, the magnetic wave vector becomes incommensurate with the lattice periodicity \cite{Pratt}.

\begin{figure}[h!]
\centering
\includegraphics[width=0.96\linewidth]{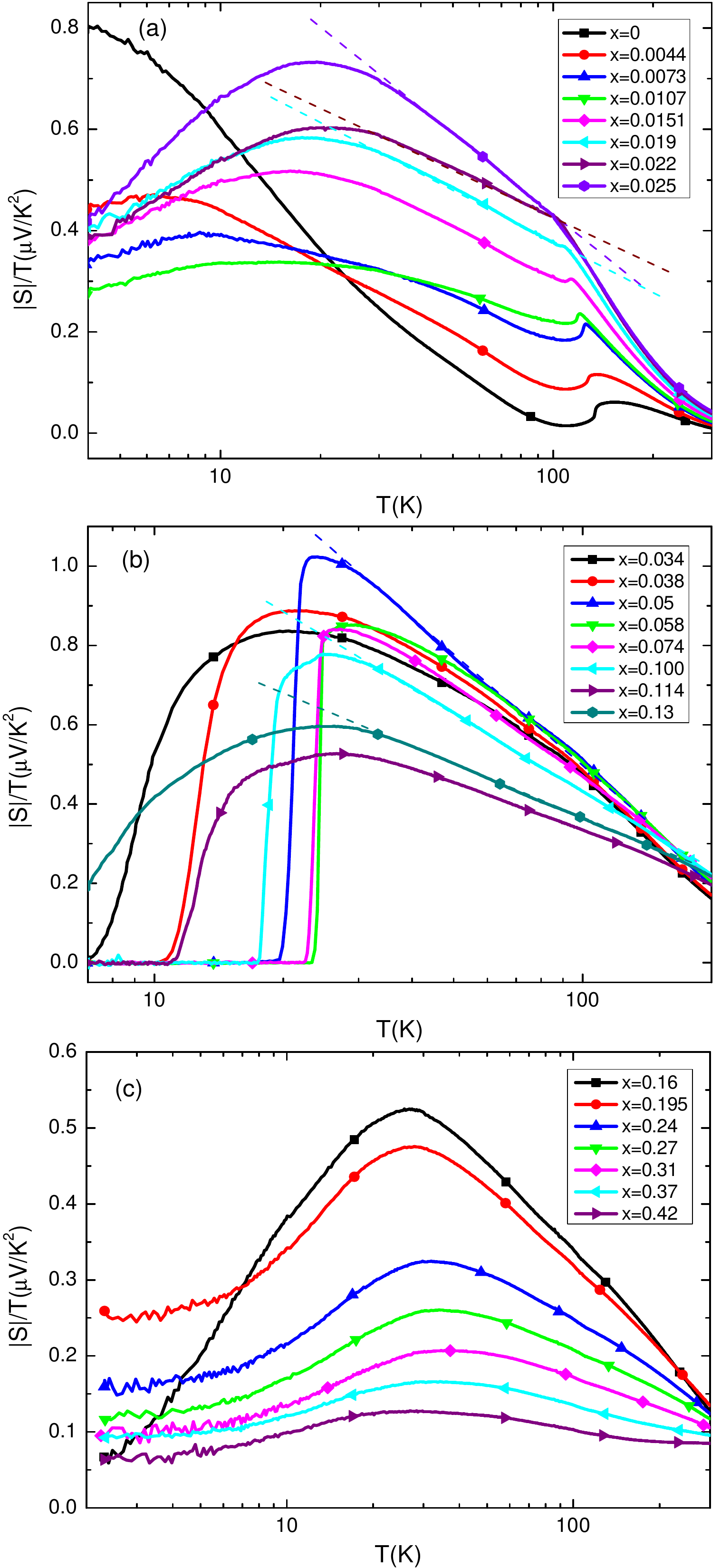}
\caption{(Color online) The behavior of $S/T$ \emph{vs} $\ln T$ in three Co substitution regimes: (a) SDW phase in which the Fermi surface is reconstructed, (b) superconducting, and (c) toward the Fermi liquid at high $x$. The dotted lines emphasize linearity on a $\log$ $T$ scale. Thermopower data are taken from \cite{Hodovanets} and \cite{Mun}.}
\label{fig:BFCA_SdT}
\end{figure}

The tightly spaced Co substitution in Ba(Fe$_{1-x}$Co$_{x}$)$_{2}$As$_{2}$ single crystals allows us to precisely map the whole $S/T$ phase diagram \cite{Hodovanets,Mun}. Thus, we can study the evolution of $S$ as the system undergoes several Lifshitz transitions \cite{BFCAFStopo}. They were observed by angle-resolved photoemission spectroscopy (ARPES)~\cite{BFCALifshitz} and by the change in thermopower and the Hall effect~\cite{Mun}. Between the first two Lifshitz transitions, interband scattering is responsible for the AF SF and thermopower is sensitive to them. Therefore, we can probe the phase diagram of Fe pnicitides in order to search for the signatures of spin-fluctuation-driven quantum criticality in $S/T$.

The temperature dependence of $|S|/T$ \emph{vs.} $\ln T$ for the whole phase diagram is presented in Fig. \ref{fig:BFCA_SdT}.
The Co concentrations used here were determined by using wavelength dispersive x-ray spectroscopy \cite{Hodovanets}. We separated the data into three groups, with each group showing a characteristic $T$ dependence.
In the first group at $x = 0-0.025$, $S/T$ undergoes an abrupt change at $T_{S}$ followed by Fermi-surface reconstruction \cite{BFCALifshitz} [Fig. \ref{fig:BFCA_SdT}(a)]. The reconstructed hole Dirac-like band in the SDW state was predicted \cite{RanDirac} and observed \cite{RichardDirac} and it induces a positive contribution to the otherwise small and compensated thermopower of $\rm BaFe_2As_2$ \cite{Arsenijevic}. This contribution to $S$ is $T$ dependent \cite{Morinari} and it is suppressed with Co substitution \cite{Hodovanets}. Its decrease is responsible for the increase of $|S|/T$ with lowering temperature in the low electron-doping regime. As we approach the Lifshitz transition at $x\approx 0.025$, the quantum critical behavior $S/T \propto \ln(1/T)$ can be observed in a limited $T$ range (30-100K).

\begin{figure}[tb]
\centering
\includegraphics[width=1.0\linewidth]{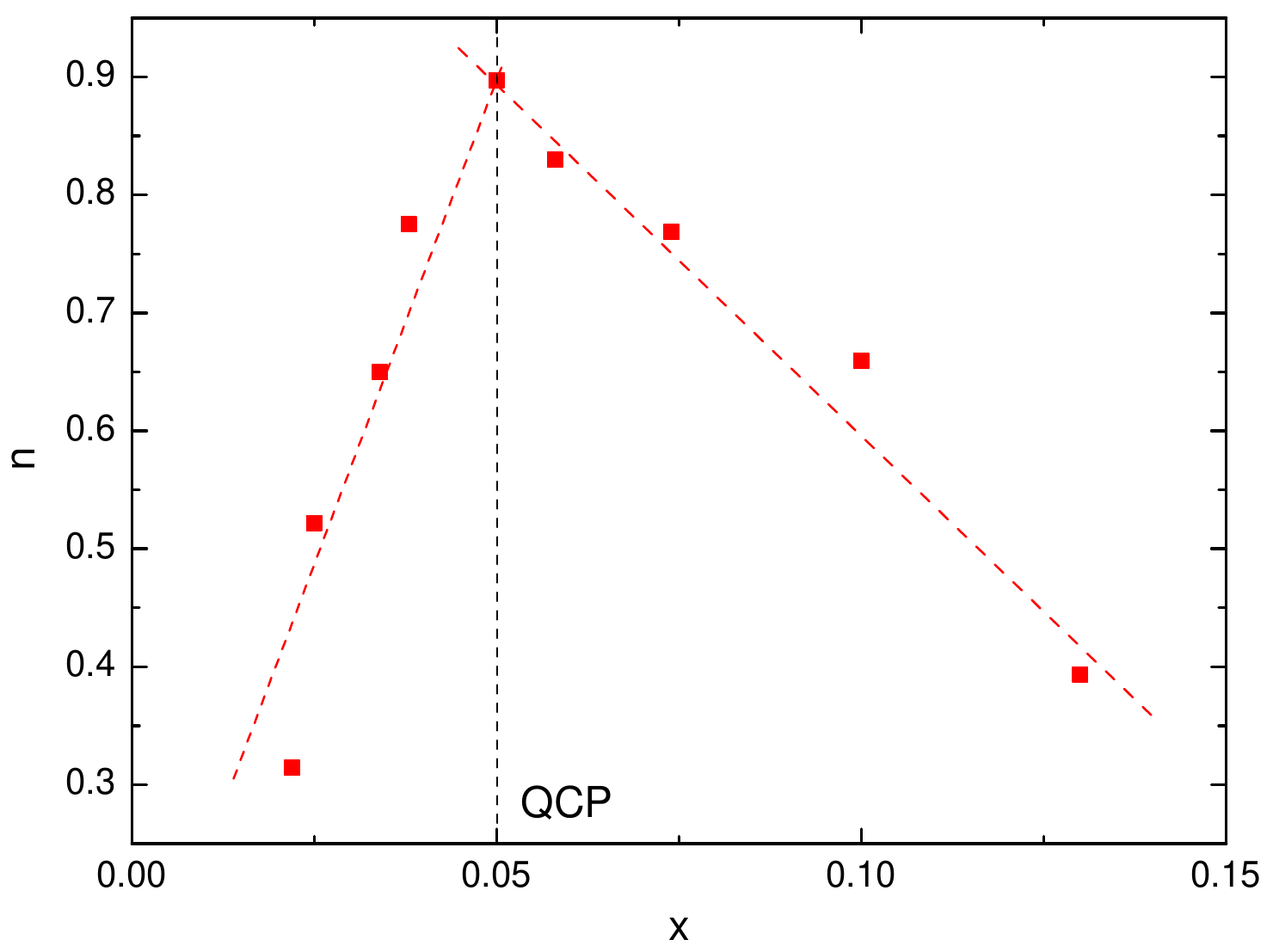}
\caption{(Color online) Plot of the slope $n$ of the logarithmic temperature dependence of $S/T$ as a function of Co substitution $x$ shows an increase close to the quantum critical point.}
\label{fig:nexp}
\end{figure}

In the second group at $x = 0.034-0.114$, the system is superconducting which is concomitant with an increase of thermopower in a large $T$ range [Fig. \ref{fig:BFCA_SdT}(b)]. The Lifshitz transition LT1 is crossed and the $S/T$ increases linearly on the $\log T$ scale with lowering $T$. With an increase of $x$, the slope of $S/T$ logarithmic $T$ dependence $n$ increases up to $x =$ 0.05 and then decreases, as shown in Fig. \ref{fig:nexp}. This can be ascribed to the decrease of Fermi energy close to the QCP, according to expression (\ref{eq:SdT}). Recent measurements of the London penetration depth on the isovalently substituted FeSC imply a decreasing effective Fermi temperature when the QCP is approached in FeSC \cite{HashimotoQCP}. Furthermore, taking into account expressions (\ref{eq:SdT}) for $S$ in the quantum critical regime and the mass of spin fluctuation $\delta = \Gamma(p - p_c) + T$, we can explain the logarithmic increase of $S$ with lowering $T$ and decreasing SF mass $\delta$. When $T < \Gamma(p - p_c)$, $S$ starts to saturate depending on the value of parameter $p$, in this case Co substitution~$x$. Above the second Lifshitz transition (LT2) at $x\approx$ 0.11 the cylindrical hole band changes to ellipsoid \cite{BFCAFStopo}, which reduces the nesting and $S/T$. In the third group, as the superconductivity is suppressed above $x =$ 0.14, the slope of $S(T)/T$ continues to decrease. This can be explained by the continuous increase of $\delta$ that results in a smaller $S$ [Fig. \ref{fig:BFCA_SdT}(c)]. The system undergoes a third Lifshitz transition (LT3) around $x\approx$ 0.2, above which the hole band is suppressed below the Fermi level. Above $x\approx$ 0.2, low-$T$ $S/T$ saturates as the system makes a cross-over from a quantum critical NFL to the Fermi-liquid-like ($S/T \propto$ const) state.

\begin{figure}[tb]
\centering
\includegraphics[width=1.0\linewidth]{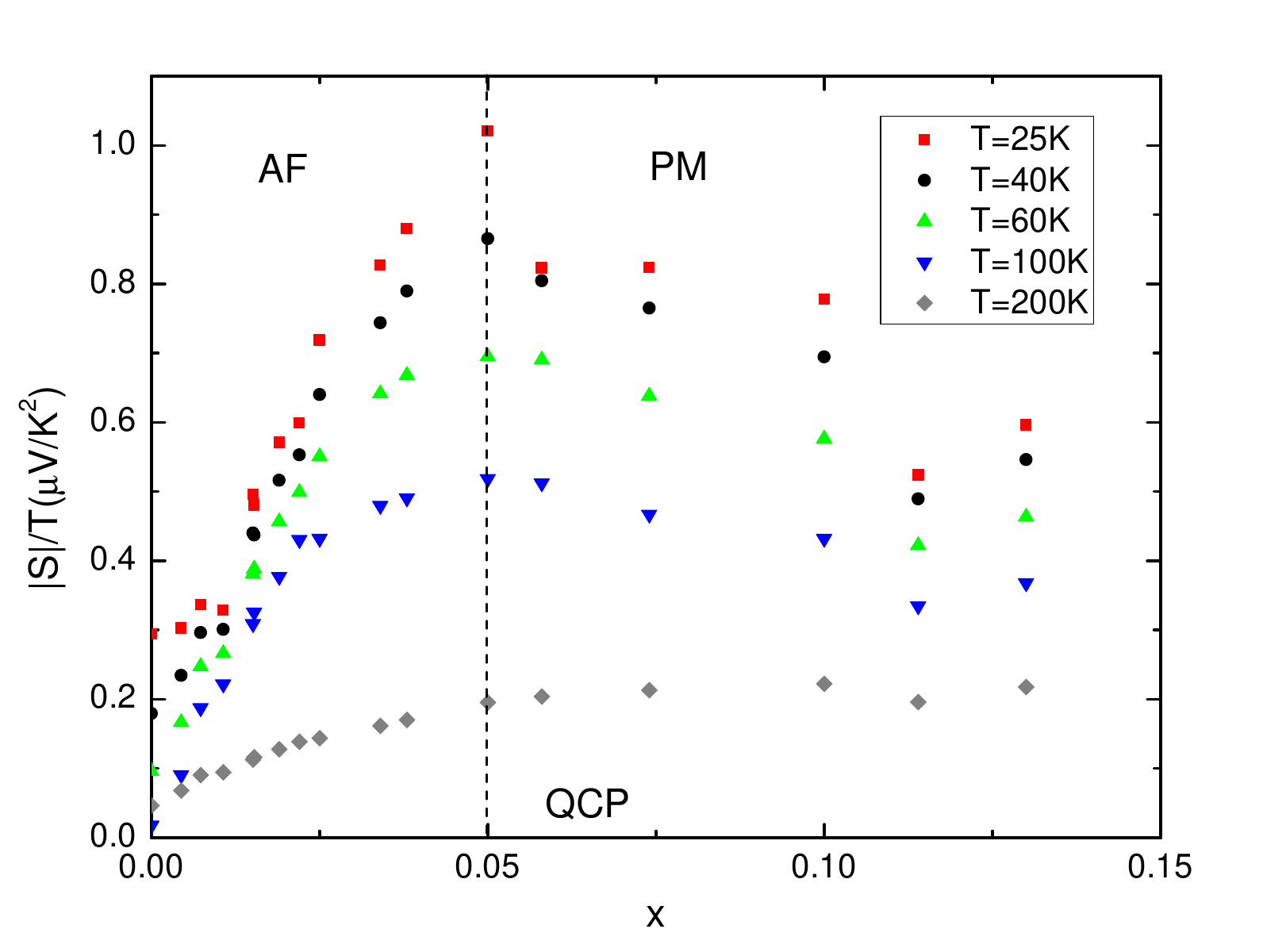}
\caption {(Color online) $S/T$ increases when approaching the QCP around $x_c \approx$ 0.05, and the parameter $\delta$ (the mass of SF) is decreasing simultaneously with a decrease of $E_F$. The increase of $S/T$ is achieved by decreasing $\delta$ with lowering temperature. In the AF phase the drop is more abrupt, similar to theoretical calculations \cite{KimPepin}.}
\label{fig:QCP_SdTx}
\end{figure}

If we analyze the $x$-dependent behavior of $S$ at fixed temperatures (Fig. \ref{fig:QCP_SdTx}), we observe that for the critical value of $x_c \approx$ 0.05, the $S/T$ attains its highest value and has a broad maximum centered at the QCP. The $x$ dependence comes from the change in the spin fluctuation mass $\delta$ and the Fermi energy in the expression for $S$. This behavior is in agreement with the theoretical calculations, which show that $S/T$ increases in proximity to the QCP in SDW systems \cite{PaulKotliar,KimPepin}. As predicted, the rate of change of $S/T(x)$ in the AF phase is more pronounced than in the paramagnetic phase, because of the reduction of entropy in the AF ordered phase. In the overdoped case ($x > 0.2$), the hole band is suppressed below the Fermi level and the bandwidth of the electron band is much larger, in agreement with the Fermi-liquid dependence seen in resistivity ($\rho \propto T^2$) \cite{FangRh,DoironPRB} and thermopower [Fig. \ref{fig:BFCA_SdT}(c)]. Opposite to that, the region closer to QCP is characterized by $T$-linear NFL $\rho$, in analogy with cuprates and Bechgaard salts \cite{DoironPRB}. A similar cross-over was observed in a heavy-fermion compound YbRh$_2$(Si$_{1-x}$Ge$_x$)$_2$, at the transition from the magnetic field-induced FL ($C_e/T \propto$ const) and the NFL state \cite{Custers}. In the same compound, $S/T$ was found to increase similarly to $C_e/T$ in the NFL state, indicating a large entropy of charge carriers. \cite{Hartmann}

\begin{figure}[tb]
\centering
\includegraphics[width=1.0\linewidth]{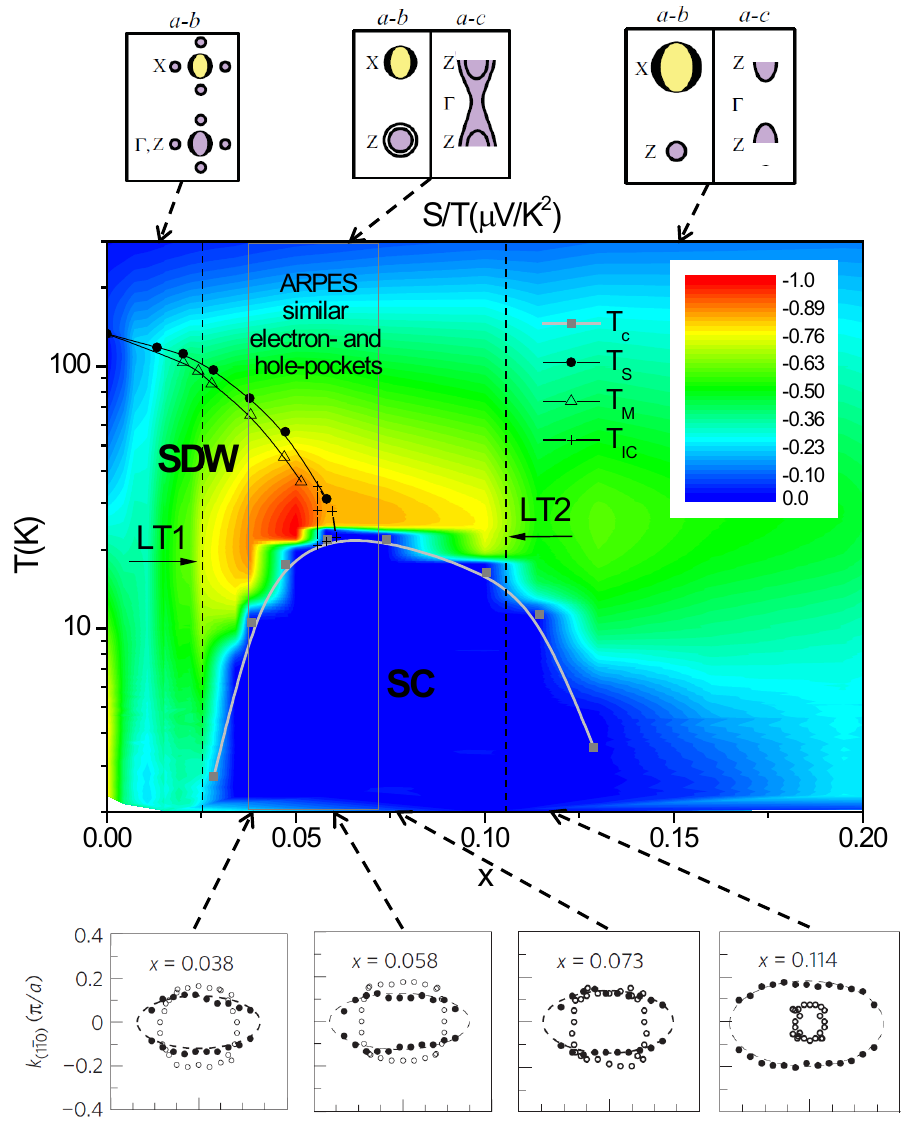}
\caption {(Color online) The contour plot of $S/T$ as a function of $\log T$ and Co substitution $x$ shows a low-$T$ increase due to spin fluctuations in the region of similar non-reconstructed electron $X$ and hole $\Gamma$ pockets, in the $x$ range between the first (LT1) and second (LT2) Lifshitz transitions. The top insets represent the scheme of Fermi-surface topology for each region in the phase diagram delimited by the Lifshitz transitions (taken from Ref. \cite{BFCAFStopo}). The lower inset emphasizes the similarity between the translated electron and hole pockets (solid- and open symbols, respectively), which can lead to the hot regions at the FS (as reported in Ref. \cite{BFCALifshitz}).
The temperatures of the superconducting ($T_c$), structural ($T_S$), and antiferromagnetic ($T_M$) transitions are taken from Refs. \cite{Ni,ChuPD}. The region of the incommensurate SDW is indicated by $T_{\rm IC}$ \cite{Pratt}.}
%The $x$-levels that correspond to the Lifshitz transitions are taken from Ref. \cite{BFCAFStopo} and the similarity between the electron- and hole-pocket was reported in Ref. \cite{BFCALifshitz}.}
\label{fig:BFCA_SdTx}
\end{figure}

A more suggestive representation of the thermopower data presented in Fig. \ref{fig:BFCA_SdT} is the contour plot of $S/T$ as a function of Co substitution and $\log T$ in Fig. \ref{fig:BFCA_SdTx}. $S/T$ maps a hot region above the SC dome, bordered by the two Lifshitz transitions, LT1 and LT2, with a dome-like distribution of intensity. In this substitution region ($x = 0.025-0.1$) the size of the electron $X$ and hole $\Gamma$ pocket is similar \cite{BFCALifshitz}. The peak of intensity is close to $x =$ 0.05, where the QCP is approached, close to the reported incommensurate spin-density-wave (IC-SDW) region, which was observed by neutron diffraction in the range 0.056 $< x <$ 0.06 \cite{Pratt}. The substitution-induced suppression of the structural and magnetic transitions coincides with the weakening of a nesting-driven SDW order, which results in an enhancement of the spin fluctuations in the region marked by the red color in Fig. \ref{fig:BFCA_SdTx}. It is the same region of the phase diagram where the back-bending of the separate SDW and structural tetragonal-to-orthorhombic transition occurs below SC $T_c$ \cite{Nandi,Pratt09}.

The observed relationship between the superconductivity, magnetism, and orthorhombicity can be explained by the magnetoelastic coupling and the closely related Ising nematic order \cite{FernandesElastic,FernandesNematic}. The electronic nematic phase with the broken $C_4$ symmetry was detected below the temperature $T^*$ by the magnetic torque and the elastic response of resistivity anisotropy measurements in the isovalently substituted system BaFe$_2$(As${1-x}$P$_x$)$2$ \cite{KasaharaNematic,Chunematic}. This phase exists above $T_S$ and is coupled to the lattice in the normal state. The nematic transition can induce the structural transition followed by the magnetic transition at a lower temperature \cite{FernandesNematic,Eremin,FernandesElastic}. The nematic instability itself is driven by the anisotropic spectrum of spin fluctuations above the AF ordered phase \cite{Eremin}. As suggested in Ref. \cite{FernandesNematic}, the scattering of electrons by SFs around the hot spots of the Fermi surface is anisotropic below the nematic transition due to the fluctuations around one of the two possible ordering vectors, ($\pi$,0) and (0,$\pi$), which become stronger than the SFs around the other vector. This leads to the anisotropic scattering of electrons and the increased in-plane resistivity anisotropy observed in Ref. \cite{ChuAniso}. In clean systems, the scattering on hot spots of the FS is hidden by the contribution from other parts of the FS \cite{HlubinaRice}. However, when the impurities are present, only electrons near the hot spots are strongly scattered by SFs, inducing a NFL behavior \cite{RoschQCP}. This effect is observable in the behavior of $S/T$ close to QCP \cite{PaulKotliar} in the BFCA system.
Approaching the QCP from the overdoped side increases the quantum spin fluctuations and the $S/T$ (Fig. \ref{fig:BFCA_SdTx}).
The region of a low-$T$ increase of thermopower in the overdoped regime is similar to the $x$ dependence of the nematic phase transition temperature $T^*$ in the paramagnetic phase of the isovalently substituted FeSC \cite{KasaharaNematic,Chunematic}.
Below the structural/magnetic transition, the spin fluctuations related to the magnetically ordered phase are indeed anisotropic and cause an anisotropic scattering \cite{FernandesNematic,Eremin}.

The link between the increase in $S/T$ and anisotropic spin fluctuations close to QCP is observed in other systems, too.
The increase in $S/T$ at $x_c$ from both the higher and lower Co substitution sides is reminiscent of the behavior observed in Sr$_3$Ru$_2$O$_7$, in which the magnetic field was used as a tuning parameter to approach the QCP \cite{RostEntropy}. Jumps in magnitude observed there in two thermodynamic variables, entropy and specific heat, were ascribed to the formation of a spin nematic phase of electronic fluid with broken rotational symmetry. This phase was previously detected as a region with highly anisotropic magnetoresistivity~\cite{Borzi}. The behavior of BFCA is analogous with this: An increase in $S/T$ in the $x$-$T$ phase diagram matches the region of increased in-plane resistivity anisotropy observed in Ref. \cite{ChuAniso}. These two phase diagrams indicate the formation of a different quantum phase, the electronic nematic phase in the vicinity of the QCP in an Fe-based superconductor, in agreement with the nematic order scenario~\cite{FernandesNematic}. This scenario is supported by the measurements of thermopower anisotropy on detwinned samples of another FeSC EuFe$_2$(As${1-x}$P$_x$)$_2$ \cite{Jiang}. An alternative scenario considers the spontaneous orbital ordering that causes the structural transition and removes the frustration of the magnetic phase, that occurs at lower temperature \cite{Krugerorbital,Chenorbital,Yinorbital}.

There are many complex systems in which IC-SDW, nematic stripe order, high thermopower and superconductivity are reported to coexist. For example, spin entropy was suggested to be responsible for the enhanced $S$ in Na$_x$Co$_2$O$_4$ (Ref. \cite{WangEntropy}) and superconducting Na$_x$Co$_2$O$_4$$\cdot\gamma H_2O$ \cite{Takada,FisherNaxCoH}. One can argue that this can be generalized to other complex transition-metal oxides, including the high-$T_c$ cuprates \cite{HinkovNematic,LaliberteStripe}.
Essentially the same behavior, compared to FeSC, was observed by the application of pressure or chemical doping to the itinerant antiferromagnet Cr \cite{Jaramillo,ChromiumSDW}. There, the nesting-driven SDW transition is suppressed with the change of external parameters, resulting in quantum critical behavior at low-$T$. Unlike FeSC, the SC $T_c$ is never high in Cr alloys because of the lack of a sufficiently attractive electron-electron interaction necessary for the Cooper pair formation \cite{ChromiumSDW}.

\section{\emph{S/T} -- Quantum criticality and superconductivity}

\begin{figure}[tb]
\centering
\includegraphics[width=1.0\linewidth]{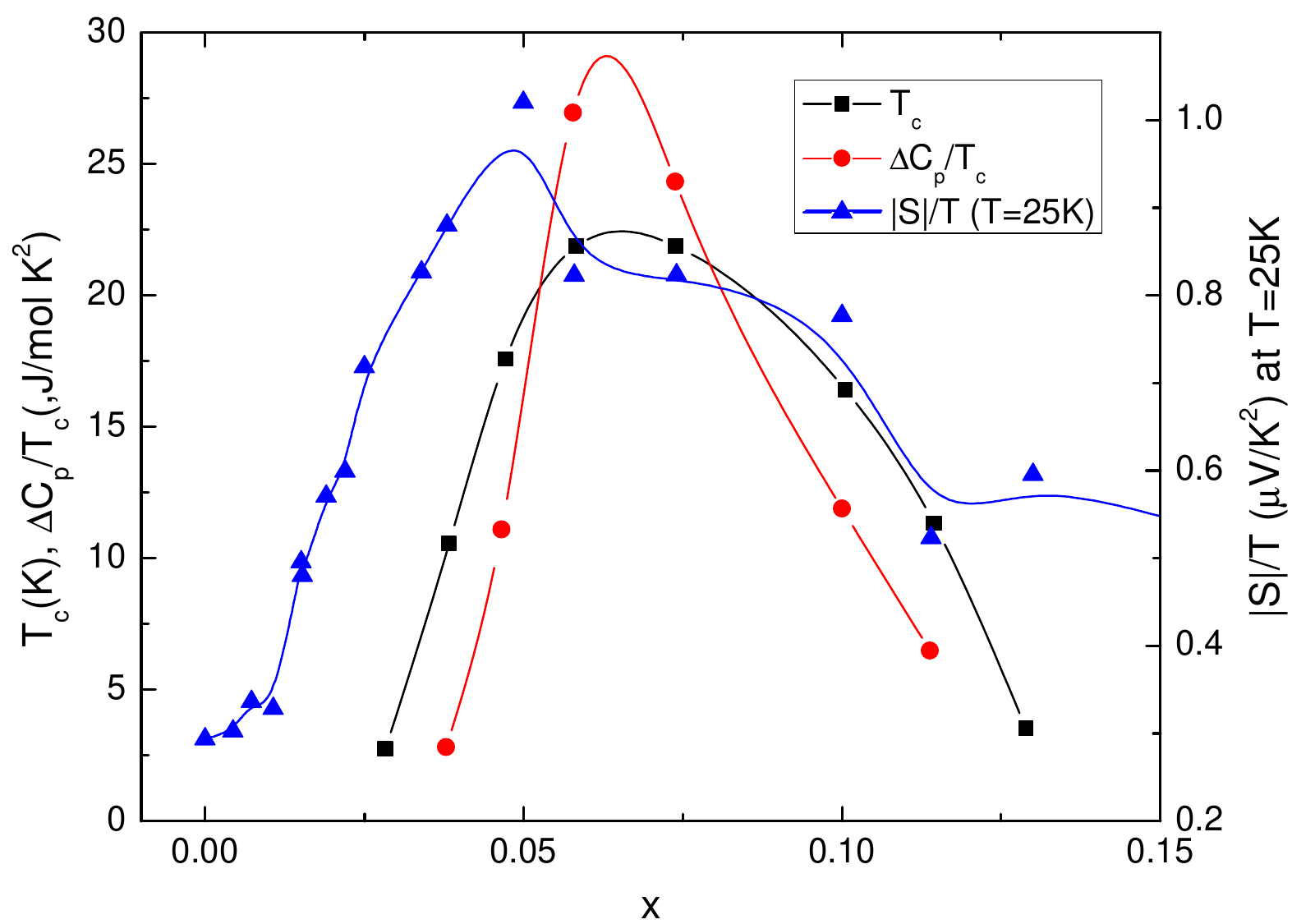}
\caption{(Color online) Superconducting transition $T_c$, specific heat jump $C_p/T_c$, and thermopower ($S/T$) at $T$ = 25 K as a function of concentration $x$ in Ba(Fe$_{1-x}$Co$_x$)$_2$As$_2$. Specific heat data are from Ref. \cite{BFCASheat}. Lines serve as a guide to the eye.}
\label{fig:BFCA_Tcdome}
\end{figure}

Strong evidence for the connection between SC and the observed quantum criticality is the correlation between the $x$ dependence of $T_c$, $S/T$, and the specific heat jump ($\Delta C_p/T_c$) at the SC $T_c$, which changes by a factor of $\sim$ 10 across the SC dome \cite{BFCASheat}.
$\Delta C_p/T_c$ \emph{vs} $T_c$ data for several FeSC can be scaled linearly to a single $\log$-$\log$ plot over an order of magnitude in $T_c$. We propose that spin entropy plays a crucial role in the maximum of $\Delta C_p/T_c$, and that the highest entropy comes from the IC-SDW for $x \sim$ 0.05. The maximum of $\Delta C_p/T_c$ corresponds to a minimum of the anisotropy of thermal conductivity and the superconducting gap modulation \cite{Reid}.
The striking similarity between the $x$ dependence of SC $T_c$, $\Delta C_p/T_c$, and $S/T$ is presented in Fig. \ref{fig:BFCA_Tcdome}. The proportionality of the $T_c$ and the strength of spin fluctuations observed in $S/T$ support the picture of SF mediated superconductivity. The SF determine the energy scale which results in the dome-like behavior in $T_c$, $\Delta C_p/T_c$, and $S/T$. This can also be observed in proximity to the thermopower intensity peak to the maximal $T_c$ in Fig. \ref{fig:BFCA_SdTx}.

The spin fluctuations are also proportional to the resonance observed in inelastic neutron scattering at the interband scattering vector \cite{Christianson}. Also, a recent, more detailed neutron study on FeSC proved that the commensuration in the spin excitation spectrum and the so-called hour-glass dispersion forms well above SC $T_c$~\cite{Tsyrulin}. The same technique detected spin excitations in the SC hole-doped Ba$_{1-x}$K$_x$Fe$_2$As$_2$, where the correlation between the Fermi surface nesting, SF energy and SC $T_c$ is observed \cite{Castellan}. The additional correlation with the critical fluctuations observed by $S/T$ in the same compound supports the argument of this paper \cite{Gooch}. A Nernst effect study on a similar compound Eu(Fe$_{1-x}$Co$_x$)$_2$As$_2$, showed the existence of an anomalous contribution that peaks above $T_c$ (around 40 K) in the sample where SDW and SC coexist \cite{Matusiak}. The authors there associated this contribution with the Fermi-surface reconstruction caused by spin fluctuations. Future Nernst effect measurements in the BFCA compound can bring useful information concerning the existence of SF above $T_c$.

\section{Conclusion}

We observe a signature of quantum critical behavior in the $T$ dependence of thermopower of the Fe-based superconductor Ba(Fe$_{1-x}$Co$_x$)$_2$As$_2$. We ascribe the increase seen in $S/T(T,x)$ around the critical substitution level $x_c$ to spin fluctuations close to the QCP. The increase of $S/T$ originates from the SDW- driven critical SFs that are enhanced at low $T$ for 0.02 $< x <$ 0.1, between the two Lifshitz transitions. In this $x$ range the electron and hole pockets are well nested, which leads to the enhanced scattering of electrons with the critical SFs at the hot regions of the Fermi surface. The smallest mass of SFs and the largest $S/T$ at low $T$ correspond to $x_c \approx 0.05$, close to the reported IC-SDW region. The quantum critical behavior that we observe in $S/T$ confirms the behavior found in $\rho$ and its anisotropy. Thus, the enhancement of thermopower and consequently, the entropy of the electron system in Fe pnictides can be related to SF, which exist above the SC $T_c$. Their strength is proportional to the $T_c$, which supports the picture of spin fluctuation mediated superconductivity.

\begin{acknowledgments}
We thank M. Sigrist, A. J\'{a}nossy, H. R{\o}nnow, I. Eremin, K. Behnia, and T. Iye for useful discussions. Work performed at EPFL was supported by the Swiss NSF and by the MaNEP NCCR. Part of this work was performed at the Ames Laboratory and supported by the U.S. Department of Energy, Office of Basic Energy Science, Division of Materials Sciences and Engineering. Ames Laboratory is operated for the U.S. Department of Energy by Iowa State University under Contract No. DE-AC02-07CH11358.
\end{acknowledgments}

\end{document}